\documentclass[12pt,preprint]{aastex}
\usepackage{natbib}

\begin{document}

\title{A search for point sources of EeV photons}

\author{The Pierre Auger Collaboration\altaffilmark{98}} 
\altaffiltext{98}{\url{www.auger.org.ar}; \url{www.auger.org}}
\altaffiltext{98}{Pierre Auger Collaboration, Av. San Mart\'{\i}n Norte 306, 5613 Malarg\"ue, Mendoza, Argentina}

\author{
\begin{small}
A.~Aab$^{42}$, 
P.~Abreu$^{65}$, 
M.~Aglietta$^{54}$, 
M.~Ahlers$^{95}$, 
E.J.~Ahn$^{83}$, 
I.~Al Samarai$^{29}$, 
I.F.M.~Albuquerque$^{17}$, 
I.~Allekotte$^{1}$, 
J.~Allen$^{87}$, 
P.~Allison$^{89}$
% ...
\\ \vspace{0.5cm}
% ...
$^{1}$ Centro At\'{o}mico Bariloche and Instituto Balseiro (CNEA-UNCuyo-CONICET), San 
Carlos de Bariloche, 
Argentina \\
% ...
$^{17}$ Universidade de S\~{a}o Paulo, Instituto de F\'{\i}sica, S\~{a}o Paulo, SP, 
Brazil \\
% ...
$^{29}$ Institut de Physique Nucl\'{e}aire d'Orsay (IPNO), Universit\'{e} Paris 11, CNRS-IN2P3, 
Orsay, 
France \\
% ...
$^{42}$ Universit\"{a}t Siegen, Siegen, 
Germany \\
% ...
$^{54}$ Osservatorio Astrofisico di Torino  (INAF), Universit\`{a} di Torino and Sezione INFN, 
Torino, 
Italy \\
% ...
$^{65}$ Laborat\'{o}rio de Instrumenta\c{c}\~{a}o e F\'{\i}sica Experimental de Part\'{\i}culas - LIP and  
Instituto Superior T\'{e}cnico - IST, Universidade de Lisboa - UL, 
Portugal \\
% ...
$^{83}$ Fermilab, Batavia, IL, 
USA \\
% ...
$^{87}$ New York University, New York, NY, 
USA \\
% ...
$^{89}$ Ohio State University, Columbus, OH, 
USA \\
% ...
$^{95}$ University of Wisconsin, Madison, WI, 
USA \\
% ...
(\ddag) Deceased \\
(a) Now at Konan University \\
(b) Also at the Universidad Autonoma de Chiapas on leave of absence from Cinvestav \\
(c) Now at NYU Abu Dhabi \\
\end{small}
}

\begin{abstract}
Measurements of air showers made using the hybrid technique developed with the fluorescence and surface detectors of the Pierre Auger Observatory allow a
sensitive search for point sources of EeV photons anywhere in the
exposed sky.  A multivariate analysis reduces the background of
hadronic cosmic rays.  The search is sensitive to a declination band from $-85^\circ$ to
$+20^\circ$, in an energy range from $10^{17.3}$~eV
to $10^{18.5}$~eV. No photon point source has been detected.  
An upper limit on the photon flux has been derived for every direction. The mean value of the energy flux limit that results from this,
assuming a photon spectral index of $-2$, is 0.06~eV~cm$^{-2}$~s$^{-1}$, and no celestial direction exceeds 0.25~eV~cm$^{-2}$~s$^{-1}$.
These upper limits constrain scenarios in which EeV cosmic ray protons are emitted by non-transient sources in
the Galaxy.
\end{abstract}


\section{Introduction}

A direct way to identify the origins of cosmic rays is to find fluxes
of photons (gamma rays) coming from discrete sources. This method has been used to
identify several likely sources of cosmic rays up to about 100 TeV
in the Galaxy~\citep{Ackermann:2013wqa}. At sufficiently
high energies, such photons must be produced primarily by $\pi^0$
decays, implying the existence of high-energy hadrons that cause the
production of $\pi^0$ mesons at or near the source. It is not known whether the Galaxy produces
cosmic rays at EeV energies (1 EeV = $10^{18}$ eV).  An argument in
favor is that the ``ankle'' of the cosmic-ray energy spectrum near 5~EeV is the
only concave upward feature, and the transition from a Galactic
power-law behavior to an extragalactic contribution should be
recognizable as just such a spectral hardening~\citep{Hillas1984}. The ankle can be explained alternatively as a ``dip'' due to $e^\pm$ pair production in a cosmic-ray spectrum that is dominated by protons of extragalactic origin. In that case, detectable sources of EeV protons would not be expected in the Galaxy \citep{Berezinsky:2004fk}.

Protons are known to constitute at least a significant fraction of the cosmic rays near the ankle of the energy spectrum \citep{Collaboration:2012wt,Abraham:2010yv,Abreu:2013env}. Those protons are able to produce photons with energies near 1 EeV by pion photoproduction or inelastic nuclear collisions near their sources. A source within the Galaxy could then be identified by a flux of photons arriving from a single direction.

The search here is for fluxes of photons with energies from
$10^{17.3}$~eV up to $10^{18.5}$~eV. The energy range is chosen to account for high event statistics and to avoid additional shower development processes that may introduce a bias at highest energies \citep{Risse:2007sd}.

The Pierre Auger Observatory~\citep{AugerFull} has excellent
sensitivity to EeV photon fluxes due to its vast collecting area and
its ability to discriminate between photons and hadronic cosmic rays~\citep{Abraham:2009qb}.
The surface detector array (SD)~\citep{Allekotte:2007sf} consists of 1660
water-Cherenkov detectors spanning 3000 km$^2$ on a regular grid of triangular
cells with 1500~m spacing between nearest neighbor stations.  It is
located at latitude $-35.2^\circ$ in Mendoza Province, Argentina.
Besides the surface array, there are 27 telescopes of the air
fluorescence detector (FD)~\citep{Abraham:2009pm} located at five sites
on the perimeter of the array.  The FD is used to measure the longitudinal
development of air showers above the surface array. The signals in the water-Cherenkov detectors are used to obtain the secondary particle density at ground measured as a function of distance to the shower core. The analysis presented in this work uses showers
measured in hybrid mode (detected by at least one FD telescope and one
SD station). The hybrid measurement technique provides a precise geometry and energy determination with a lower energy detection threshold compared to SD only measurements~\citep{Abraham:2009pm}. Moreover, multiple characteristics of photon-induced air
showers can be exploited by the two detector systems in combination, e.g., muon-poor ground signal and large depth of shower maximum compared to hadronic cosmic rays of the same energy. Several
photon--hadron discriminating observables are defined and combined in
a multivariate analysis (MVA) to
search for photon point sources and to place directional upper limits
on the photon flux over the celestial sphere up to declination
$+20^\circ$.

The sensitivity depends on the declination of a target direction.  For
the median exposure, a flux of 0.14~photons~km$^{-2}$~yr$^{-1}$ or greater
would yield an excess of at least 5$\sigma$.  This corresponds to an
energy flux of 0.25~eV~cm$^{-2}$~s$^{-1}$ for a photon flux following a
$1/E^2$ spectrum, similar to energy fluxes of sources measured by TeV
gamma-ray detectors.  This is relevant because the energy flux per
decade is the same in each energy decade for a source with a $1/E^2$
spectrum, and Fermi acceleration leads naturally to such a type of
spectrum (cf.\ Sec.\ \ref{sec:ResultsDiscussion}).  The Auger Observatory has the sensitivity to detect photon
fluxes from such hypothetical EeV cosmic-ray sources in the Galaxy.

At EeV energies, fluxes of photons are attenuated over intergalactic
distances by $e^\pm$ pair production in collisions of those photons with
cosmic-background photons. The $e^\pm$ can again interact with background photons via inverse-Compton scattering, resulting in an electromagnetic cascade that ends at GeV-TeV energies. The detectable volume of EeV photon sources is small compared to the Greisen-Zatsepin-Kuz'min (GZK)
sphere~\citep{Greisen1966,Zatsepin1966}, but large enough to encompass
the Milky Way, the Local Group of galaxies and possibly Centaurus A,
given an attenuation length of about 4.5~Mpc at EeV energies~\citep{Risse:2007sd, DeAngelis:2013jna,EleCa}.

The present study targets all exposed celestial directions without
prejudice.  It is a ``blind'' search to see if there might be an indication
of a photon flux from any direction.  One or more directions of
significance might be identified for special follow-up study with
future data.  Because there is a multitude of ``trials,'' some excesses are likely to occur by chance. A genuine modest flux would not be detectable in this kind of blind search.  The possible
production of ultra-high energy photons and neutrons has been studied extensively in
relation to some directions in the Galaxy \citep{CiteMedina,
  Bossa:2003fa, Aharonian:2004jr, Crocker:2004bb, Gupta:2011aw}.  
   
The Auger Collaboration has published stringent upper limits on the
\textit{diffuse} intensity of photons at ultra-high
energies~\citep{Abraham:2006ar,Aglietta:2007yx,Abraham:2009qb,Abreu:2011pf}.  Those limits impose
severe constraints on ``top-down'' models for the production of
ultra-high energy cosmic rays.  At the energies in this study,
however, the limits do not preclude photon fluxes of a strength that
would be detectable from discrete directions.

The paper is organized as follows: in Sec.\ \ref{sec:AnalysisMethod}, mass composition-sensitive observables are introduced, exploiting information from the surface detector as well as from the fluorescence telescopes. These observables are combined in a multivariate analysis explained in Sec.\ \ref{sec:MVA}, before introducing the dataset and applied quality cuts in Sec.\ \ref{sec:Dataset}. A calculation of the expected isotropic background contribution is described in Sec.\ \ref{sec:background}. A blind search technique and an upper limit calculation are explained in Sec.\ \ref{sec:BlindSearch} and Sec.\ \ref{sec:UL}, respectively. Finally, results are shown and discussed in Sec.\ \ref{sec:ResultsDiscussion}.

\section{Mass composition-sensitive observables}
\label{sec:AnalysisMethod}
The strategy in searching for directional photon point sources is based on the selection of a subset of photon-like events, to reduce the isotropic hadronic background. Such a selection relies on the combination of several mass composition-sensitive parameters, using a multivariate analysis (MVA). 

Once the MVA training is defined, the photon-like event selection is optimized direction-wise, accounting for the expected background contribution from a given target direction to take into account the contribution of different trigger efficiencies. 

Profiting from the hybrid nature of the Auger Observatory, we make use of FD- and SD-based observables, which provide complementary information on the longitudinal and lateral distributions of particles in the showers, respectively.  
By means of Monte Carlo (MC) simulations, five observables are selected to optimize the signal (photon) selection efficiency against the background (hadron) rejection power. The selected observables are described below in detail. 

\subsection{FD observables}
A commonly used mass-composition sensitive observable is the depth of the shower maximum $X_{\rm max}$, which is defined as the atmospheric depth at which the longitudinal development of a shower reaches its maximum in terms of energy deposit. Given their mostly electromagnetic nature, on average, photon-induced air showers develop deeper in the atmosphere, compared to hadron-induced ones of similar energies, resulting in larger $X_{\rm max}$ values. The difference is about 100~g~cm$^{-2}$ in the energy range discussed in this paper. The reconstruction procedure is based on the fit of the Gaisser-Hillas function~\citep{GaisserHillas1977} to the energy deposit profile, which has been proven to provide a good description of the Extensive Air Shower (EAS) independently of the primary type. \\

In addition to the Gaisser-Hillas function, the possibility of fitting the longitudinal profile with the Greisen function~\citep{ref.Greisen:1965} has been explored. The Greisen function was originally introduced to describe the longitudinal profile of pure electromagnetic showers: a better fit to the longitudinal profile is thus expected for photon-initiated showers when compared to nuclear ones of the same primary energy. The $\chi_{\rm Gr}^2/{\rm dof}$ is used to quantify the goodness of the fit and as potential discriminating observable. 
The Greisen function has one free parameter, that is the primary energy $E_{\rm Gr}$, which is also influenced by the primary particle. The observable is $E_{\rm Gr} / E_{\rm GH}$, where $E_{\rm GH}$ is the energy obtained by integrating the Gaisser-Hillas function.  All of the $X_{\rm{max}}$, the $\chi_{\rm Gr}^2/{\rm dof}$ and the  $E_{\rm Gr} / E_{\rm GH}$ are used as variables contributing to the photon-hadron classifier. The method adopted for the classification, as well as the relative weight of each variable to the classification process, will be discussed in the next section. 

\subsection{SD observables}
When observed at ground, photon-induced showers have a generally steeper lateral distribution than nuclear primaries because of the almost absent muon component. It is worth noting that, as a consequence of the trigger definition in the local stations and of the station spacing in the array~\citep{SD_acceptance}, the surface detector alone is not fully efficient in the energy range used in this work. Thus, as opposed to previous work based on SD observables~\citep{Aglietta:2007yx}, we adopt here observables that are defined at the station level and which do not necessarily require an independent reconstruction in SD mode. Such observables are related to an estimator ($S_b$) of the lateral distribution of the signal or to the shape of the flash analog digital converter (FADC) trace in individual stations.

The $S_b$ parameter is sensitive to different lateral distribution functions, due to the presence/absence of the flatter muon component~\citep{ref.Ros:2011}, and has already been used in previous studies~\citep{Abreu:2011pf}. It is defined as 
\begin{equation}
 S_b = \sum_{i=1}^N \left[ S_i \cdot \left( \frac{r_i}{1000~{\rm m}}  \right)^b \right]~,
\end{equation}
where the sum extends over all $N$ triggered stations, $S_i$ expresses the signal strength of the $i$--th SD station, $r_i$ the distance of this station to the shower axis, and $b$ a variable exponent. It has been found that, in the energy region of interest, the optimized $b$ for photon--hadron separation is $b=3$~\citep{Ros:2013lxa}.\\
As a result of both the smaller signal in the stations, on average, and the steeper lateral distribution function, smaller values of $S_b$ are expected for photon primaries. 
To prevent a possible underestimate of $S_b$ (which would mimic the behavior of a photon-like event), due to missing stations during the deployment of the array or temporarily inefficient stations, events are selected requiring at least 4 active stations (fully operational, but not necessary triggered) within 2~km from the core. 
 
Other observables, containing information on the fraction of electromagnetic and muonic components at the ground, are related to measurements of the time structure derived from the FADC traces in the SD. The spread of the arrival times of shower particles at a fixed distance from the shower axis increases for smaller production heights, i.e., closer to the detector station. Consequently, a larger spread is expected in case of deep developing primaries (i.e., photons). 
Here we introduce the shape parameter, defined as the ratio of the early-arriving to the late-arriving integrated signal as a function of time measured in the water-Cherenkov detector with the strongest signal:
\begin{equation}
{\rm ShapeP}(r,\theta) = \frac{S_{\rm early}(r,\theta)}{S_{\rm late}(r,\theta)}~.
\end{equation} 
The early signal $S_{\rm early}$ is defined as the integrated signal over time bins less than a scaled time $t_i^{\rm scaled} \leq 0.6$~$\mu$s, beginning from the signal start moment. The scaled time varies for different inclination angles $\theta$ and distances $r$ to the shower axis and can be expressed as: 
\begin{equation}
t_i^{\rm scaled}(r,\theta)=t_i \cdot \frac{r_0}{r} \cdot \frac{1}{c_1 + c_2 \cdot \cos (\theta)}~,
\end{equation} 
where $t_i$ is the real time of bin $i$ and $r_0=1000$~m is a reference distance. $c_1=-0.6$ and $c_2=1.9$ are scaling parameters to average traces over different inclination angles. Correspondingly, the late signal $S_{\rm late}$ is the integrated signal over time bins later than $t_i^{\rm scaled} > 0.6$~$\mu$s, until signal end.

\section{Multivariate analysis}
\label{sec:MVA}

The selected discriminating observables are combined by a multivariate analysis technique to enhance and maximize their photon-hadron separation power. In particular, the analysis was developed by using a Boosted Decision Tree (BDT) as classifier~\citep{Breiman1984,Schapire1990}. Several other classifiers were also tested, but the BDT stands out due to the simplicity of the method, where each training step involves a one-dimensional cut optimization, in conjunction with high-performance photon--hadron discrimination~\citep{Kuempel2011}. Another advantage of BDTs is that they are insensitive to the inclusion of poorly discriminating variables. The observables were selected from a larger sample of SD and hybrid observables by looking at their individual discrimination power in different energy and zenith bins, their strength in the BDT, and the stability of the results. One benchmark quantity that assess the performance of BDTs is the separation. It is defined to be zero for identical signal and background shapes of the output response, and it is one for shapes with no overlap. The overall separation as well as the separation if excluding a specific observable from the analysis is listed in Table \ref{tab:MVA}. 
\begin{table}[h]
\begin{center}
    \caption{Overall separation of the observables using BDTs (\textit{all}) as well as the remaining separation if excluding a single observable from the MVA.}
    \begin{tabular}{ |l|l|}
    \hline
    \textit{Observables} & \textit{Separation}  \\ \hline \hline
     all & 0.668    \\ \hline
     No $S_b$ & 0.438    \\ \hline
     No $X_{\rm max}$ & 0.599    \\ \hline
     No $\chi^2_{\rm Gr} / {\rm dof}$ & 0.662    \\ \hline
     No $E_{\rm Gr}/E_{\rm GH}$ & 0.664    \\ \hline
     No ShapeP & 0.667    \\ \hline
     \end{tabular}
    \label{tab:MVA}
\end{center}
\end{table}
The most significant variables contributing to the BDT are $X_{\rm max}$ and $S_{b}$. With these variables alone we achieve a separation of 0.654. During the classification process the BDT handles the correlation of the observables to energy and zenith angle of the primary particle by including them as additional parameters.

\begin{figure}[t]
\centering
\includegraphics[width=11cm]{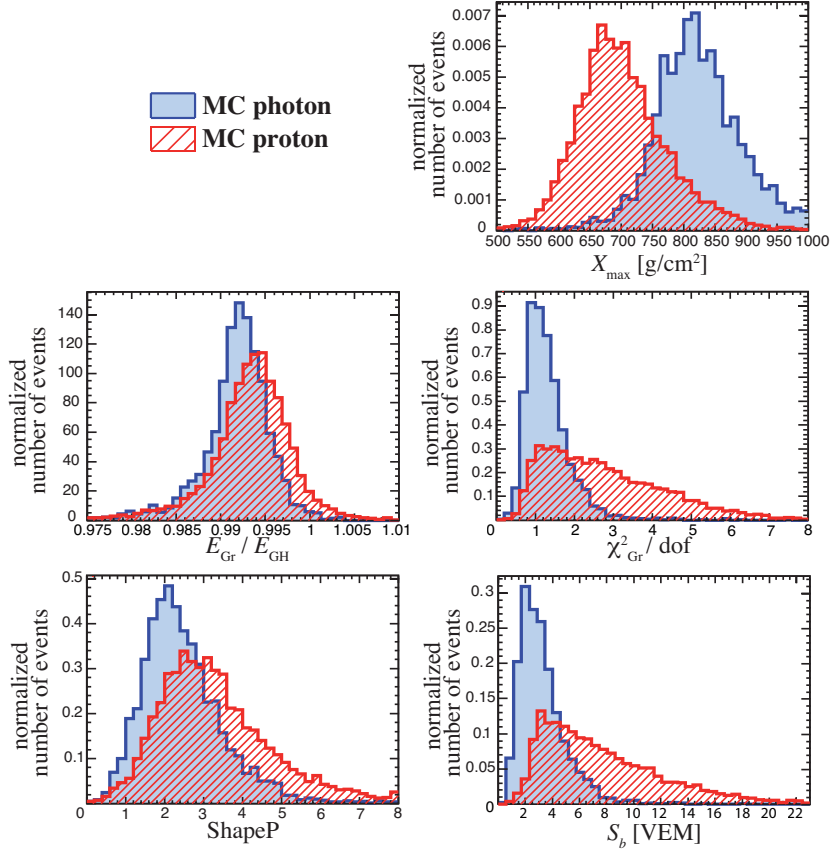}
\caption{Distributions of photon (full blue) and proton (striated red) simulations of the introduced observables. The distributions are shown as examples for the energy range between $10^{17.6}$~eV and $10^{18}$~eV and zenith angle between 0$^\circ$ and 30$^\circ$.}
\label{fig:observables}
\end{figure}

\begin{figure}[t]
\centering
\includegraphics[width=10cm]{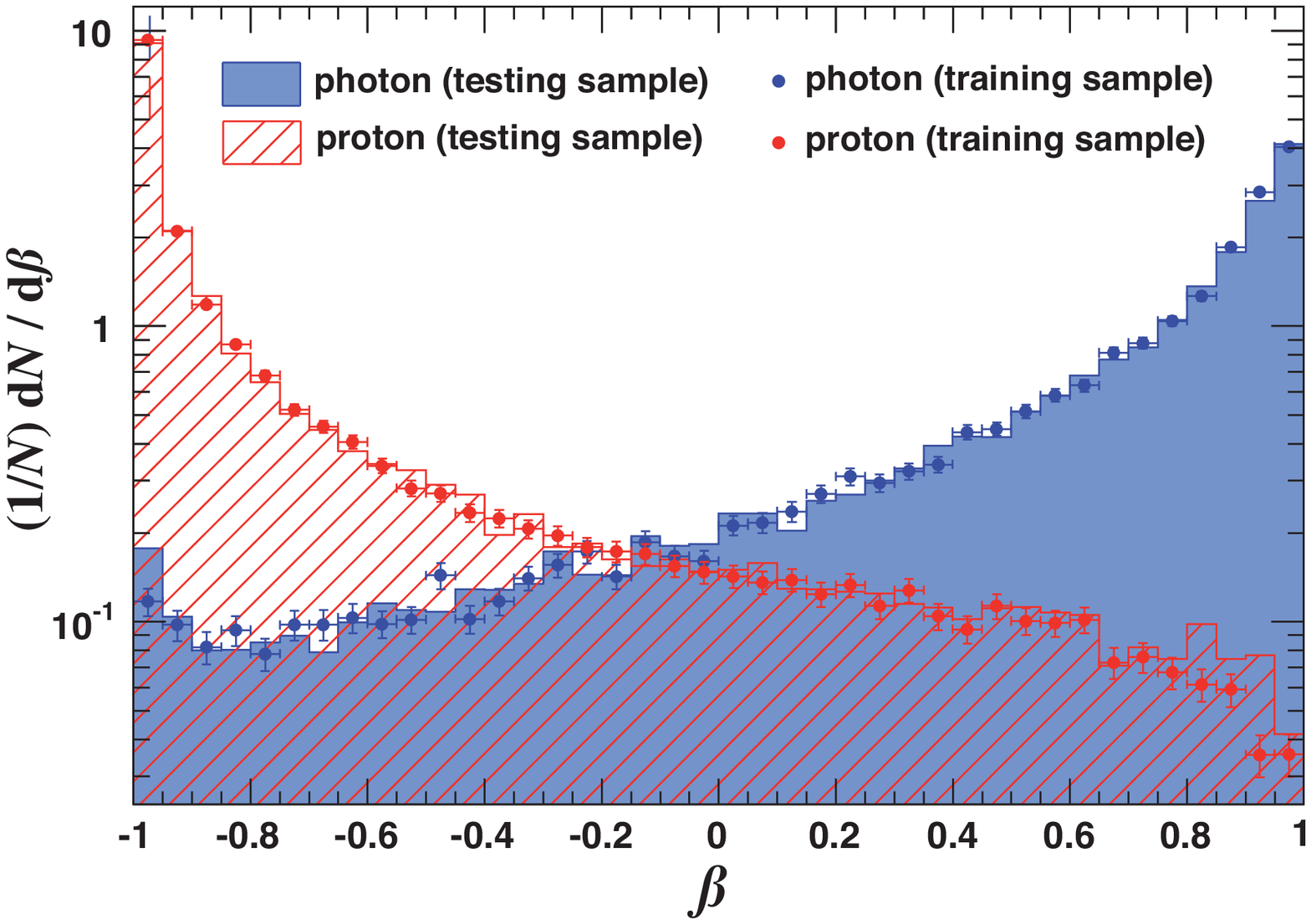}
\caption{Multivariate analysis response value $\beta$ for photon and proton primaries using boosted decision trees. During evaluation the MC sample is split half into a training (filled circles) and half into a testing sample (solid line).}
\label{fig:OutResponse}
\end{figure}

For the classification process, BDTs are trained and tested using MC simulations. Air showers are simulated using the CORSIKA v.\ 6.900~\citep{HeckCORSIKA} code. A total number of $\sim 30,000$ photon and $\sim 60,000$ proton primaries are generated according to a power-law spectrum of index -2.7 between $10^{17.2}$~eV and $10^{18.5}$~eV, using QGSJET-01c~\citep{Kalmykov:1989br} and GHEISHA~\citep{citeGHEISHA} as high- and low-energy interaction models, respectively. The impact of different hadronic interaction models is discussed in Sec.\ \ref{sec:ResultsDiscussion}.
The detector response was simulated using the simulation chain developed within the offline framework~\citep{Argiro:2007qg}, as discussed in~\citep{Settimo2012}. The same reconstruction chain and the same selection criteria as for data (discussed in Sec.\ \ref{sec:Dataset}) were then applied.
During the classification phase, photon and proton showers are reweighted according to a spectral index of -2.0 and -3.0, respectively. The impact of a changing photon spectral index on the results is discussed in Sec.\ \ref{sec:ResultsDiscussion}. The distribution of observables for photon and proton simulations for a specific energy and zenith range is shown in Fig.\ \ref{fig:observables}. 
The MVA output response value is named $\beta$ and shown in Fig.\ \ref{fig:OutResponse} for the training and testing samples. Note that the $\beta$ distribution is by construction limited to the range $[-1,1]$.

\section{Dataset}
\label{sec:Dataset}
\begin{figure}[t]
\centering
\includegraphics[width=9cm]{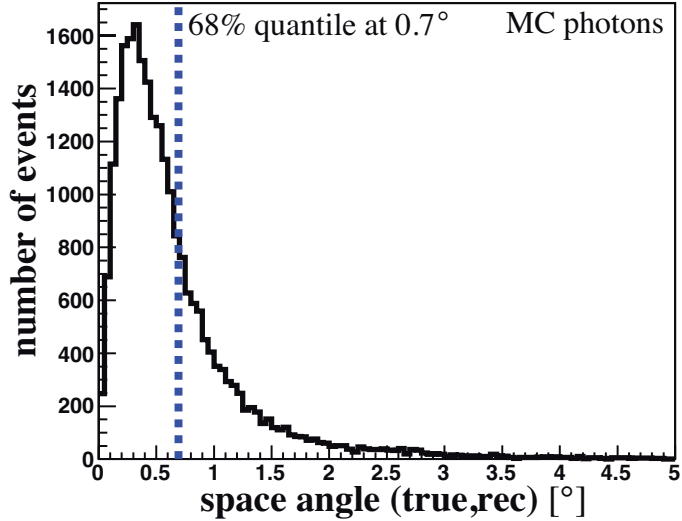}
\caption{Space angle distribution between simulated and reconstructed arrival direction of photon primaries. The angular resolution is calculated as the 68\% quantile located at 0.7$^\circ$ denoted by the dotted line.}
\label{fig:resolution}
\end{figure}

The search for photon point sources is performed on the sample of hybrid events collected between January 2005 and September 2011, under stable data-taking conditions. Events are selected requiring a reconstructed energy between $10^{17.3}$~eV and $10^{18.5}$~eV, where the energy is determined as the calorimetric one plus a 1\% missing energy correction associated with photon primaries~\citep{Barbosa:2003dc}. To ensure good energy and directional reconstruction, air showers with zenith angle smaller than $60^\circ$ and with a good reconstruction of the shower geometry are selected. For a reliable profile reconstruction we require: a reduced $\chi^2$ of the longitudinal profile fit to the Gaisser-Hillas function smaller than 2.5, a Cherenkov light contamination smaller than 50\%, and an uncertainty of the reconstructed energy less than 40\%. Overcast cloud conditions can distort the light profiles of EAS and influence also the hybrid exposure calculation~\citep{CloudPAO}. To reject misreconstructed profiles, we select only periods with a detected cloud coverage $\leq 80\%$ with a cut efficiency of 91\%. In addition, only events with a reliable measurement of the vertical optical depth of aerosols are selected \citep{BenZvi:2006xb}. As already mentioned, at least 4 active stations are required within 2~km of the hybrid-reconstructed axis to prevent an underestimation of $S_b$. To enrich our sample with deep showers, we do not require that $X_{\rm max}$ has been observed within the field of view. This cut is usually applied to assure a good $X_{\rm max}$ resolution (e.g.\ \citep{Abraham:2009qb, Abraham:2010mj, Abraham:2010yv}), but for this analysis we focus on a maximization of the acceptance for photon showers. Profiles, for which only the rising edge is observed are certain to have a deep $X_{\rm max}$ below ground level ($\sim 840/\cos(\theta)$~g/cm$^2$). Shallow showers, that are also enriched by the release of this cut, can be easily removed in later stages of the analysis by the MVA $\beta$ cut (cf.\ Sec.\ \ref{sec:BlindSearch}). The photon $X_{\rm max}$ resolution with this type of selection increases from 39 to 53 g/cm$^2$, but the photon acceptance is increased by 42\%. The energy resolution is about 20\%, independently of the primary mass. These resolutions do not affect significantly the analysis since the trace of photons from a point source is an accumulation of events from a specific direction, and the event direction is well reconstructed also with the relaxed cuts: as shown in Fig.\ \ref{fig:resolution}, the angular resolution is about 0.7$^\circ$. We also verified that the separation power of the MVA is not significantly modified by the weaker selection requirements.

After selection, the final dataset consists of $N_{\rm data} = 241,466$ events with an average energy of $10^{17.7}$~eV. In fact, the energy distribution of these events expresses a compensation effect of the energy spectral index and trigger inefficiencies at low energies. A discussion of the hybrid trigger efficiency for hadrons in the energy range below $10^{18}$~eV is given in~\citep{Settimo2012}. In Sec.\ \ref{sec:UL} this discussion is extended to the case of photons. The average number of triggered stations in the current dataset is 2 at $10^{17.5}$~eV, where the bulk of events is detected, generally increasing with zenith angle and with energy (up to 4 between $10^{18}$~eV and $10^{18.5}$~eV).

\section{Background expectation}
\label{sec:background}

The contribution of an isotropic background is estimated using the scrambling technique~\citep{Cassiday:1989yw}. This method has the advantage of using only measured data and takes naturally into account detector efficiencies and aperture features. Therefore, it is not sensitive to the (unknown) cosmic ray mass composition in the covered energy range.

As a first step, the arrival directions (in local coordinates) of the events are smeared randomly according to their individual reconstruction uncertainty. In a second step, $N_{\rm data}$ events are formed by choosing randomly a local coordinate and, independently, a Coordinated Universal Time from the pool of measured directions and times. This procedure is repeated 5,000 times. The mean number of arrival directions within a target is then used as the expected number for that particular sky location. As each telescope has a different azimuthal trigger probability, events are binned by telescope before scrambling. The number of events observed in each telescope varies between 4,358 and 14,100. Since the scrambling technique is less effective in the southern celestial pole region\footnote{At the pole, the estimated background would always be similar to the observed signal. Therefore a possible excess or deficit of cosmic rays from the pole would always be masked.}, declinations $< -85^\circ$ are omitted from the analysis. 

Sky maps are pixelized using the HEALPix software~\citep{ref.Gorski2005}. Target centers are taken as the central points of a HEALPix grid using $N_{\rm side} = 256$ (target separation $\sim 0.3^\circ$), resulting in 526,200 target centers south of a declination of $+20^\circ$.  The treatment of arrival directions is based on an unbinned analysis, i.e., angular distances are calculated analytically. 
For each target direction, we use a top-hat counting region of 1$^\circ$ (selecting events within a hard cut on angle from the target center), motivated to account for low event statistics (cf.\ \citep{Alexandreas:1992ek}) and a possible non-Gaussian tail of the error distribution\footnote{It was verified that selecting containment radii of $0.74^{\circ}$ and $1.5^\circ$ increases the mean flux upper limit of point sources by $+9\%$ and $+11\%$, respectively.}.

\begin{figure}[t]
\centering
\includegraphics[width=12cm]{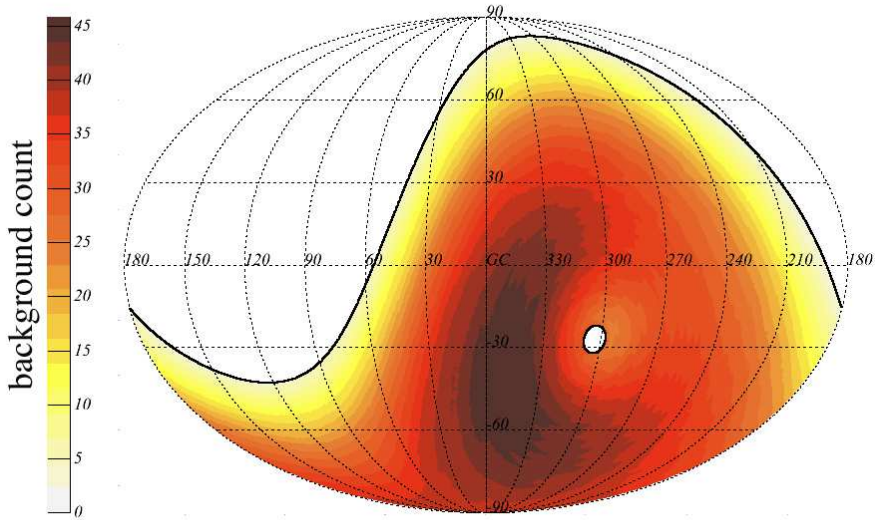}
\caption{Sky map of the expected background contribution (average of 5,000 scrambled maps) in Galactic coordinates using the Mollweide-projection~\citep{CiteMollweide}. The solid black lines indicate the covered declination range between $-85^\circ$ and +20$^\circ$. Note that the southern celestial pole region is omitted in this analysis for reasons explained in Sec.\ \ref{sec:background}.}
\label{fig.BackgroundCount}
\end{figure}

The expected directional background contribution for the covered search period is shown in Fig.\ \ref{fig.BackgroundCount}. There is an azimuthal asymmetry in the expected background as a result of a seasonally-dependent duty cycle, i.e., during austral summer, data taking using the fluorescence telescopes is reduced compared to austral winter~\citep{Settimo2012, exposure}.  

\section{Blind search analysis}
\label{sec:BlindSearch}
\begin{figure}[t!]
\centering
\includegraphics[width=12cm]{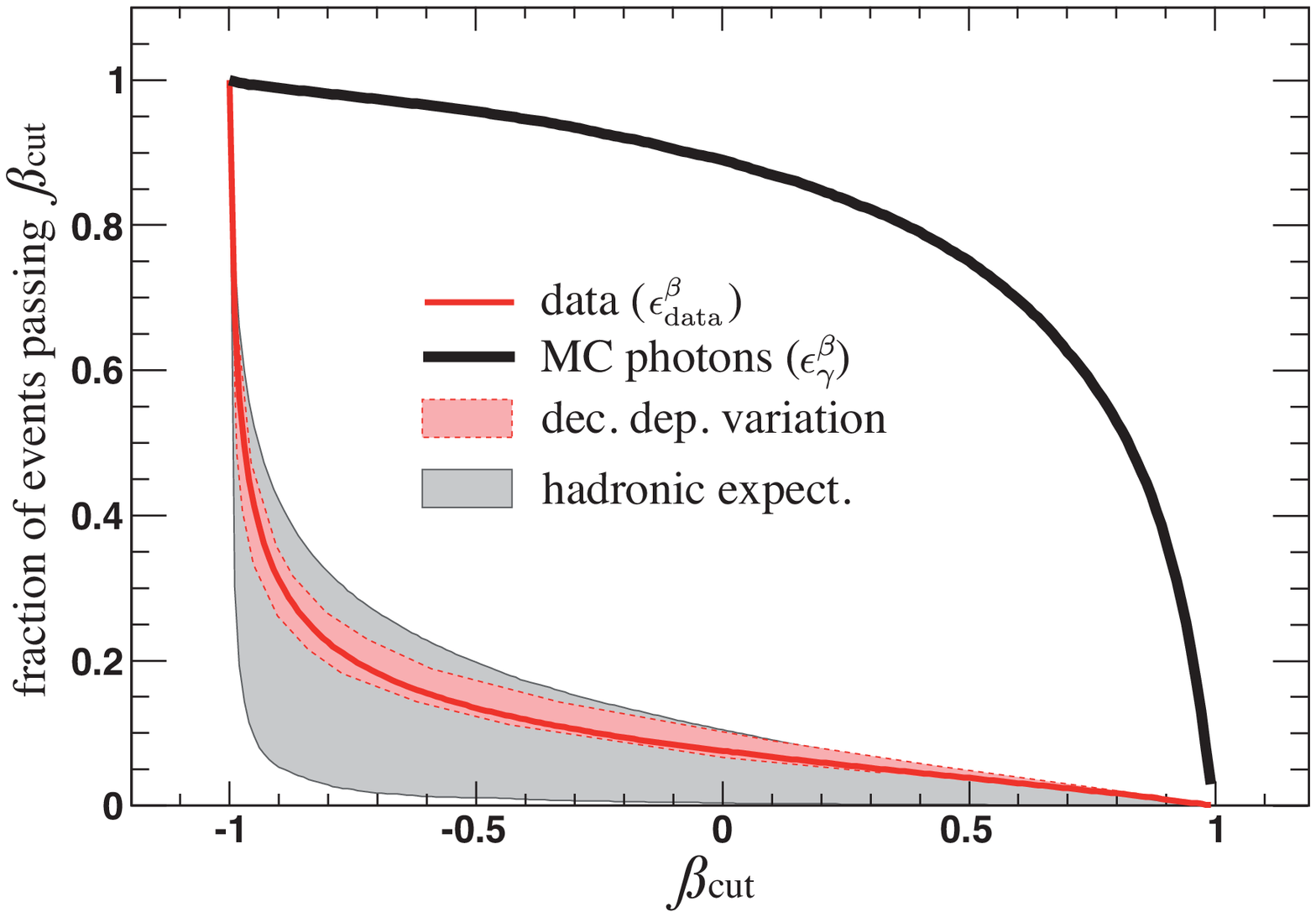}
\caption{Fraction of events passing the $\beta_{\rm cut}$ for simulated primary photons (black) and measured averaged data (red). The red shaded area represents the declination-dependent variation of the data. The grey shaded area represents the expectation of a purely hadronic composition derived from MC simulations.}
\label{fig:CutEff}
\end{figure}

\begin{figure}[t!]
\centering
\includegraphics[width=12cm]{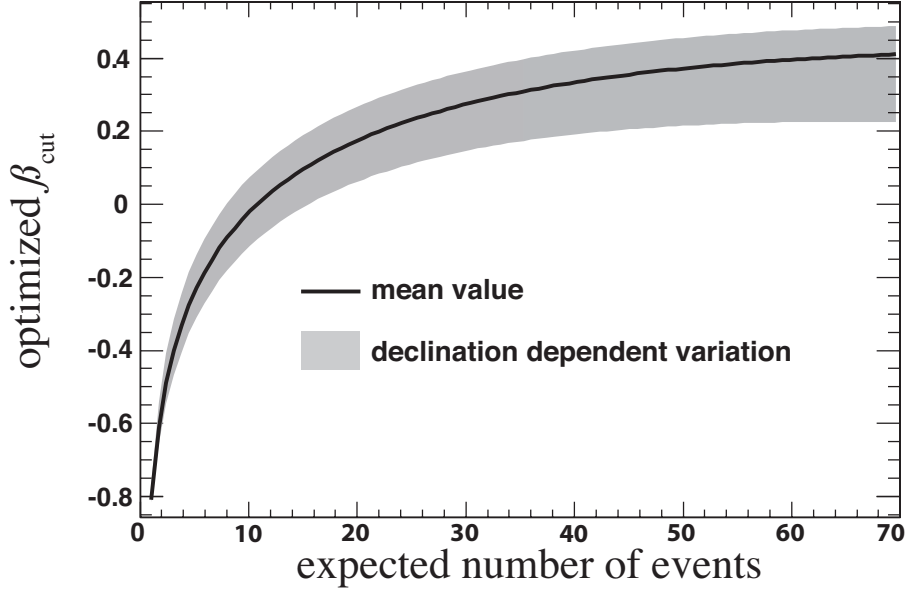}
\caption{Optimized $\beta_{\rm cut}$ as a function of the expected background count. The mean value (solid black line) and the declination-dependent variations (shaded area) are illustrated.}
\label{fig:OptCut}
\end{figure}

\begin{figure}[t]
\centering
\includegraphics[width=12cm]{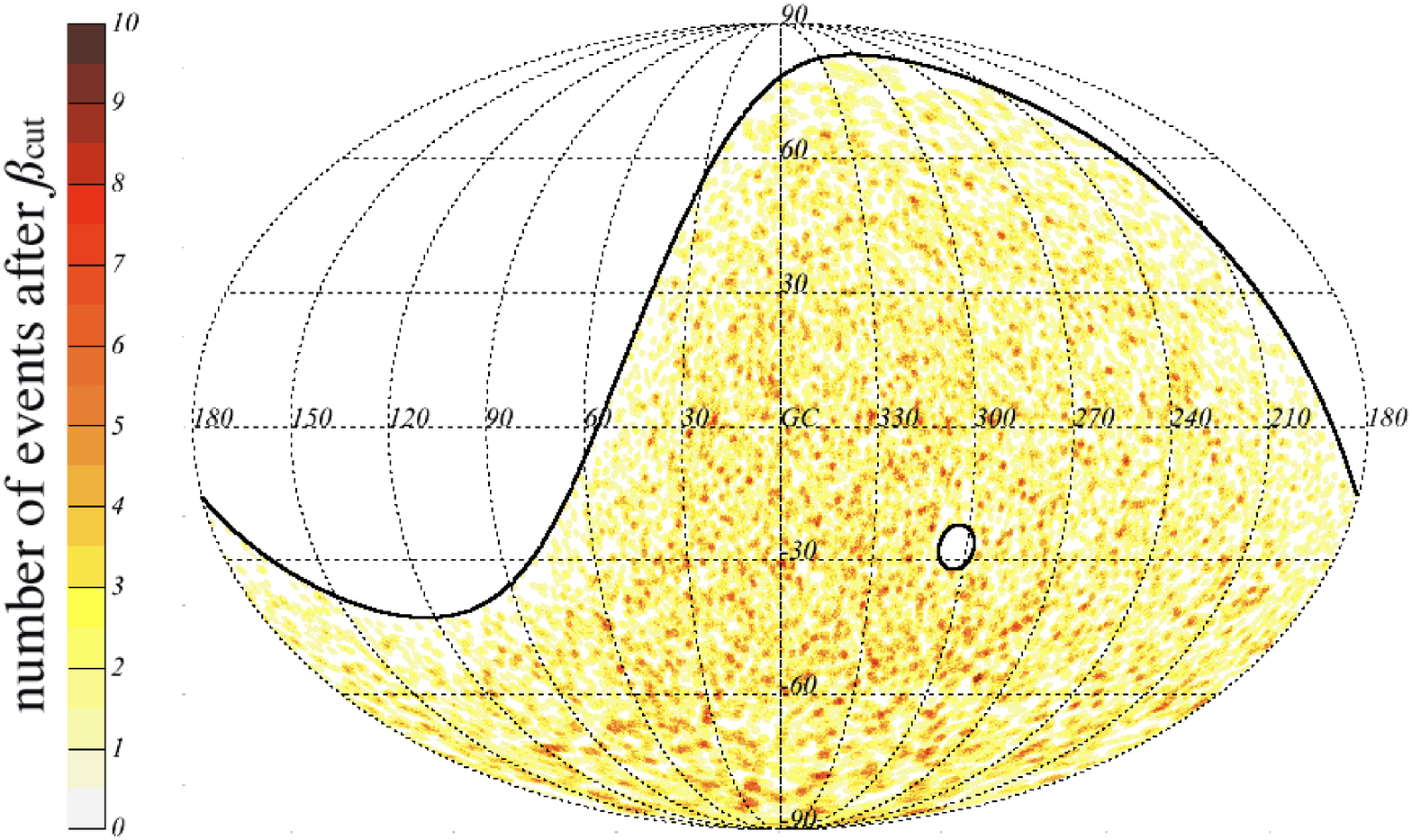}
\caption{Sky map of measured events after applying the optimized $\beta_{\rm cut}$ illustrated in galactic coordinates.}
\label{fig.EventsAfterBetaCut}
\end{figure}

When performing the blind search analysis, we use only a subset of the recorded data, selected as ``photon-like'' according to the $\beta$ distribution. The definition of ``photon-like'' (i.e., the $\beta_{\rm cut}$ position when selecting events with $\beta \ge \beta_{\rm cut}$) is related to the MC photon and the data efficiencies, $\varepsilon_\gamma^\beta$ and $\varepsilon_{\rm data}^\beta$, respectively, and to the expected number of background events $n_b(\alpha, \delta)$ which is a function of the celestial coordinates $\alpha$ and $\delta$. The efficiencies $\varepsilon_\gamma^\beta$ and $\varepsilon_{\rm data}^\beta$ are shown in Fig.\ \ref{fig:CutEff} as a function of the multivariate cut $\beta_{\rm cut}$. To estimate $\varepsilon_{\rm data}^\beta$ more accurately, a declination dependence is taken into account, $\varepsilon_{\rm data}^\beta=\varepsilon_{\rm data}^\beta ({\rm \delta})$, indicated by the red shaded area in Fig.\ \ref{fig:CutEff}. The expectation of a purely hadronic composition is shown as a grey band. 
To improve the detection potential of photons from point sources, the cut on the $\beta$ distribution is optimized, dependent on the direction of a target center or, more specifically, dependent on the expected number of background events $n_b(\alpha, \delta)$. In this way the background contamination is reduced while keeping most of the signal events in the dataset. This optimization procedure can be described as follows: the upper limit of photons $n_s$ from a point source at a given direction is calculated under the assumption that $n_{\rm data} = n_b^{\beta}$, i.e., when the observed number of events ($n_{\rm data}$) is equal to the expected number (cf.\ Sec.\ \ref{sec:background}). The expected number of events after cutting on the $\beta$ distribution can be estimated as $n_b^\beta (\alpha, \delta) = n_b(\alpha, \delta) \cdot \varepsilon_{\rm data}^\beta (\delta)$ and is typically less than 4 events for the $\beta_{\rm cut}$ values finally chosen. There are several ways to define an upper limit on the number of photons $n_s$, at a given confidence level (CL) in the presence of a Poisson-distributed background. Here the procedure of Zech~\citep{ref.Zech1989} is utilized, where $n_s$ is given by: 
\begin{equation}
P(\leq n_b^\beta | n_b^\beta + n_s) = \alpha_{\rm CL} \cdot P(\leq n_b^\beta | n_b^\beta)~,
 \label{eqn:Zech}
\end{equation}
with $\alpha_{\rm CL} \equiv 1-{\rm CL} = 0.05$, and where the expected background contribution is $n_b^\beta$ (cf.\ \citep{Auger:2012yc}). The frequentist interpretation of the above equation is as follows: ``For an infinitely large number of repeated experiments looking for a signal with expectation $n_s$ and background with mean $n_b^\beta$, where the background contribution is restricted to a value less than or equal to $n_b^\beta$, the frequency of observing $n_b^\beta$ or fewer events is $\alpha_{\rm CL}$.'' Since $n_b^\beta$ is not an integer in general, a linear interpolation is applied to calculate the Poisson expectation. To determine the optimized $\beta_{\rm cut}$, the sensitivity is maximized by minimizing the expected upper limit by scanning over the entire range of possible $\beta_{\rm cut}$, also taking into account the photon efficiency $\varepsilon_\gamma^\beta$:
\begin{equation}
\min \left( \frac{n_s(\beta_{\rm cut})}{\varepsilon_\gamma^\beta (\beta_{\rm cut})} \right)~~~{\rm with}~\beta_{\rm cut} \in [-1,1].
\end{equation}
The optimized mean $\beta_{\rm cut}$ is shown in Fig.\ \ref{fig:OptCut} as a function of the expected background contribution. The grey area indicates the declination-dependent variation of the optimization.  
The mean $\beta_{\rm cut}$ value used in this analysis is 0.22 resulting in an average background contribution after $\beta_{\rm cut}$ of 1.48 events. Applying the optimized $\beta_{\rm cut}$ to measured data reduces the dataset to 13,304 events. The sky distribution of these events is shown in Fig.\ \ref{fig.EventsAfterBetaCut}.

When performing a {blind search for photon point sources, the probability $p$ of obtaining a test statistic at least as extreme as the one that was actually observed is calculated, assuming an isotropic distribution. The test statistic is obtained from the ensemble of scrambled datasets (cf.\ Sec.\ \ref{sec:background}), assuming a Poisson-distributed background. This $p$-value is calculated for a specific target direction as:
\begin{equation}
p = {\rm Poiss} (\geq n_{\rm data}^\beta | n_b^\beta )~,
\label{eqn:pValue}
\end{equation}
where ${\rm Poiss} (\geq n_{\rm data}^\beta | n_b^\beta )$ is the Poisson probability to observe $n_{\rm data}^\beta$ or more events given a background expectation after $\beta_{\rm cut}$ of $n_b^\beta$. Note that the superscript ``$\beta$'' indicates the number of events after applying the optimized $\beta_{\rm cut}$. The fraction of simulated datasets $p_{\rm chance}$, in which the observed minimum $p$-value $p_{\rm min}$ is larger than or equal to the simulated $p$-value $p_{\rm min}^{\rm scr}$, is given by:
\begin{equation}
p_{\rm chance}(p_{\rm min}^{\rm scr} \leq p_{\rm min})~.
\label{eqn:pChance}
\end{equation}
This corresponds to the chance probability of observing $p_{\rm min}$ anywhere in the sky. The results when applying this blind search to the hybrid data of the Pierre Auger Observatory will be discussed in Sec.\ \ref{sec:ResultsDiscussion}.

\section{Upper limit calculation}
\label{sec:UL}

Here we specify the method used to derive a skymap of upper limits to the photon flux of point sources. The directional upper limit on the photon flux from a point source is the limit on the number of photons from a given direction, divided by the directional acceptance (cf.\ Sec.\ \ref{sec:background}) from the same target at a confidence level of ${\rm CL} = 95\%$, and by a correction term:
\begin{equation}
f^{\rm UL} = \frac{n_s^{\rm Zech}}{n_{\rm inc} \cdot \mathcal{E}_{\rm{\beta}}}~.
\label{eqn.FluxUpperLimit}
\end{equation}
Here $n_s^{\rm Zech}$ is the upper limit on the number of photons obtained by using the $\beta_{\rm{cut}}$ definition in Fig.~\ref{fig:OptCut}, and applying the procedure of Zech (cf.\ Eqn.\ (\ref{eqn:Zech})) for the observed number of events in data $n_{\rm data}^\beta$:
\begin{equation}
 P(\leq n_{\rm data}^\beta | n_b^\beta + n_s^{\rm Zech}) = \alpha_{\rm CL} \cdot P(\leq n_{\rm data}^\beta | n_b^\beta)~.
\end{equation}
The expected signal fraction in the top-hat search region is $n_{\rm inc}=0.9$, and $\mathcal{E_\beta}$ is the total photon exposure. This latter exposure is derived as:
\begin{equation}
\mathcal{E}_{\beta}(\alpha, \delta) = \mathcal{E}(\alpha, \delta)\cdot \varepsilon_\gamma^\beta~, 
\end{equation}
where $\mathcal{E}$ indicates the exposure before applying the multivariate cut $\beta_{\rm{cut}}$ (cf.\ Eqn.\ \ref{eqn:exp}), and $\varepsilon_\gamma^\beta$  is the photon efficiency when applying a $\beta_{\rm{cut}}$.

The exposure $\mathcal{E}(E)$ is typically defined as a function of energy $E$, cf.\ \citep{exposure,Settimo2012,Abraham:2010mj}. In a similar way, the photon exposure $\mathcal{E}$ as a function of celestial coordinates $\alpha$ and $\delta$ is defined as: 
\begin{equation}
\mathcal{E}(\alpha, \delta) = \frac{1}{c_E} \int_{E}\int_T\int_{S}  E^\zeta~\varepsilon(E,t,\theta,\phi,x,y) ~\textrm{d}S ~ \textrm{d}t ~ \textrm{d}E~,
\label{eqn:exp}
\end{equation}
where the coordinates $\alpha$ and  $\delta$ are functions of the zenith ($\theta$) and azimuth ($\phi$) angles and of the time $t$; $\varepsilon$ is the overall efficiency including detection, reconstruction and selection of the events and the evolution of the detector in the time period $T$. The integration over energy is performed assuming a power-law spectrum with index $\zeta = -2$ and normalization factor $c_E = \int E^{\zeta}~{\rm d}E$. The area $S$ encloses the full detector array and is chosen sufficiently large to ensure a negligible (less than 1\%) trigger efficiency outside of it. The exposure for the hybrid detector is not constant with energy and is not uniform in right ascension. Thus, detailed simulations were performed to take into account the status of the detector and the dependence of its performance with energy and direction (both zenith and azimuth). For the exposure calculation applied here, time-dependent simulations were performed, following the approach described in~\citep{exposure,Settimo2012}. This takes into account the photon trigger efficiency, possible periods of overcast cloud conditions, and offers the possibility of also investigating systematic uncertainties below the EeV range. At the low energy edge of $10^{17.3}$~eV, the hybrid trigger efficiency for photon-induced air showers is larger than 80\%, rapidly increasing before reaching full efficiency at $10^{17.8}$~eV. Compared to hadron induced air showers, the photon trigger efficiency is always larger as a consequence of a later shower development, on average. The derived directional photon exposure has a mean of 180~km$^2$~yr and varies between 50~km$^2$~yr and 294~km$^2$~yr. The impact of different photon spectral indices is discussed in Sec.\ \ref{sec:ResultsDiscussion}. Directional upper limits on photons from point sources derived in this blind search analysis will be also discussed in Sec.\ \ref{sec:ResultsDiscussion}.

\section{Results and discussion}
\label{sec:ResultsDiscussion}

\begin{figure}[t]
\centering
\includegraphics[width=10cm]{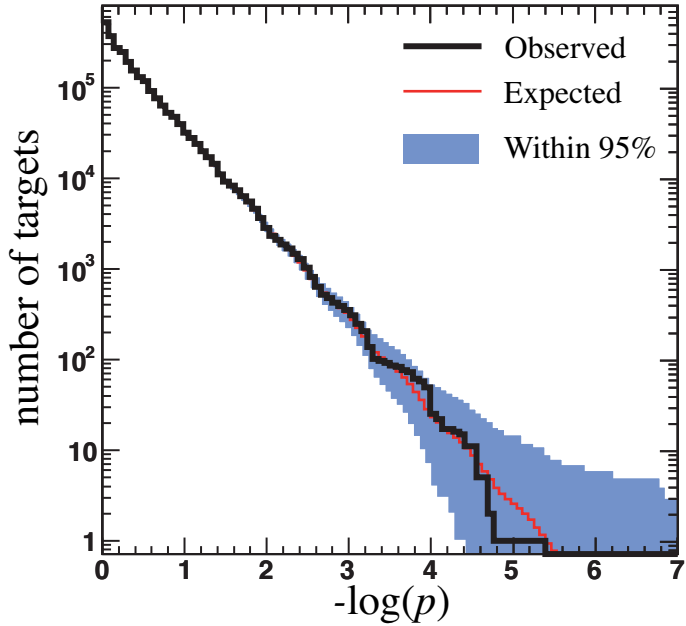}
\caption{Integral distribution of $p$-values. For better visibility $-\log(p)$ is shown. The observed distribution is shown as a thick black line, the mean expected one, assuming background only, as a thin red line. The blue shaded region corresponds to 95\% containment of simulated data sets.}
\label{fig.HistpValue}
\end{figure}

\begin{figure}[t]
\centering
\includegraphics[width=12cm]{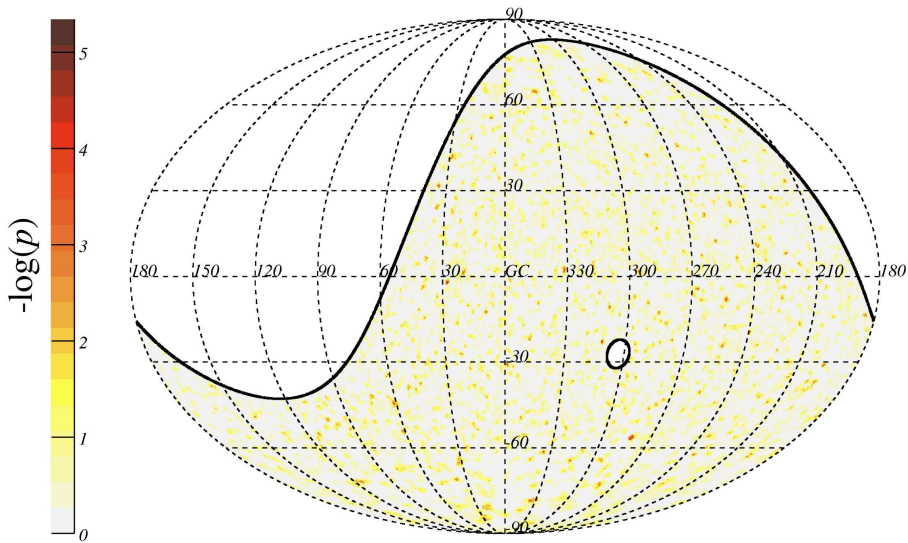}
\caption{Celestial map of $-\log(p)$ values in Galactic coordinates.}
\label{fig.SkyPValue}
\end{figure}

\begin{figure}[t]
\centering
\includegraphics[width=12cm]{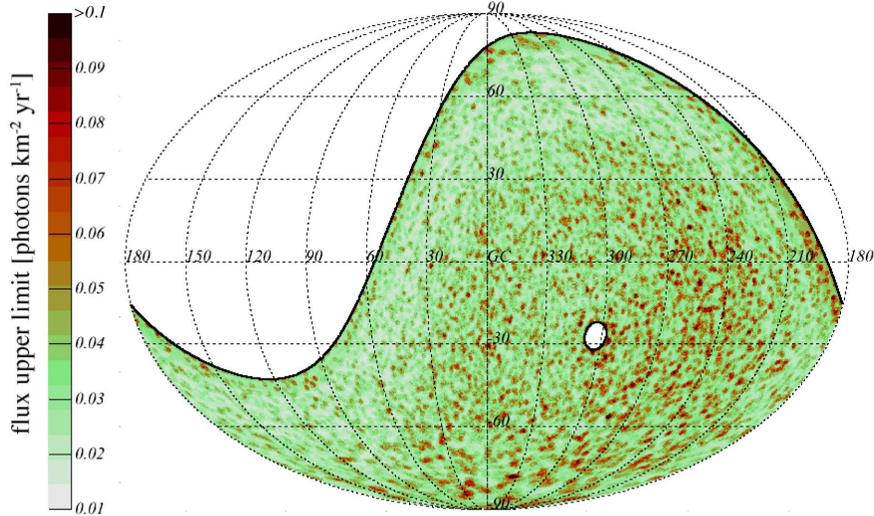}
\caption{Celestial map of photon flux upper limits in photons~km$^{-2}$~yr$^{-1}$ illustrated in Galactic coordinates.}
\label{fig.SkyFluxUL}
\end{figure}

In the following, results on $p$-values and upper limits are given, based on the analysis method described in Sec.\ \ref{sec:BlindSearch} and Sec.\ \ref{sec:UL}.

The $p$-values, as defined in Eqn.\ \ref{eqn:pValue}, refer to a local probability that the data is in agreement with a uniform distribution. The integral distribution of $-\log(p)$-values is shown in
Fig.\ \ref{fig.HistpValue}. The corresponding sky map of
$-\log(p)$-values is illustrated in Fig.\ \ref{fig.SkyPValue}. The
minimum $p$-value observed is $p_{\rm min}=4.5\times10^{-6}$
corresponding to a chance probability that $p_{\rm min}$ is observed
anywhere in the sky of $p_{\rm chance} = 36\%$.  This blind search for
a flux of photons, using hybrid data of the Pierre Auger
Observatory, therefore finds no candidate point on the pixelized
sky that stands out among the large number of trials.  It is possible that some genuine photon fluxes are responsible for some of
the low $p$-values.  If so, additional exposure should increase the
significance of those excesses.  They might also be identified in a
future search targeting a limited number of astrophysical candidates.
The present search, however, finds no statistical evidence for any
photon flux.

Directional photon flux upper limits (95\% confidence
level) are derived using Eqn.\ (\ref{eqn.FluxUpperLimit}) and shown as
a celestial map in Fig.\ \ref{fig.SkyFluxUL}. The mean value is
0.035~photons~km$^{-2}$~yr$^{-1}$, with a maximum of
0.14~photons~km$^{-2}$~yr$^{-1}$.  Those values correspond to an energy flux of
0.06~eV~cm$^{-2}$~s$^{-1}$ and 0.25~eV~cm$^{-2}$~s$^{-1}$,
respectively, assuming an $E^{-2}$ energy spectrum.

Various sources of systematic uncertainties were investigated and
their impact on the mean flux upper limit is estimated. The systematics on the photon exposure ranges between $\pm 30$\% at $10^{17.3}$~eV and $\pm 10$\% above $10^{18}$~eV and are dominated by the uncertainty on the Auger energy scale. A systematic uncertainty of the Auger energy scale of $+14\%$ and $-14\%$ \citep{ValerioICRC13} changes the mean upper limit by about $+8\%$ and $-9\%$, respectively. Variations in determining the fraction of photon ($\epsilon^\beta_\gamma$) and
measured events ($\epsilon^\beta_{\rm data}$) passing a $\beta_{\rm
  cut}$, introduced by, e.g., an additional directional dependency for
photons, contribute less than $6\%$. A
collection of $\sim 50,000$ proton CORSIKA air shower simulations,
using EPOS LHC~\citep{Pierog:2013ria}, were additionally generated
to estimate the impact of using a different high-energy hadronic
interaction model. The resulting change of the mean limit of the
photon flux is $-9\%$. Furthermore, the assumed photon flux spectral
index of $-2$ could be incorrect. To estimate the impact of this, the analysis is
repeated assuming a spectral index of $-1.5$ or $-2.5$. The mean
upper limit changes by about $-34\%$ and $+51\%$, respectively,
whereas the dominant contribution arises from a changing directional
photon exposure, i.e., assuming a flatter primary photon spectrum
increases the photon exposure, while reducing the average upper limits,
and vice versa.

The limits are of considerable astrophysical interest in all parts of the exposed sky. The energy
flux in TeV gamma rays exceeds 1~eV~cm$^{-2}$~s$^{-1}$ for some
Galactic sources with a differential spectral index of
$E^{-2}$~\citep{Hinton:2009zz,HESSCite}.  A source with a differential
spectral index of $E^{-2}$ puts out equal energy in each decade, resulting
in an expected energy flux of 1~eV~cm$^{-2}$~s$^{-1}$ in the EeV
decade. No energy flux that strong in EeV photons is observed from
any target direction, including directions of TeV sources such as Centaurus A or the Galactic center region. This flux would have been detected with $>5\sigma$ significance, even after penalizing for the large number of trials (using Eqn.\ \ref{eqn:pValue} and Eqn.\ \ref{eqn:pChance}). Furthermore, an energy flux of 0.25~eV~cm$^{-2}$~s$^{-1}$ would yield an excess of at least 5$\sigma$ for median exposure targets. If we make the conservative assumption that all detected photons are at the upper energy bound, a flux of 1.44~eV~cm$^{-2}$~s$^{-1}$ would be detectable. This result for median exposure targets is independent of the assumed photon spectral index, and implies that we can exclude a photon flux greater than 1.44~eV~cm$^{-2}$~s$^{-1}$ with 5$\sigma$ significance.

Results from the present study complement the blind search for fluxes of
{\it neutrons} above 1 EeV previously published by the Auger
Collaboration~\citep{Auger:2012yc}.  No detectable flux was found in
that search, and upper limits were derived for all
directions south of declination $+20^\circ$.  A future study will look for evidence of photon fluxes from particular
candidate sources and ``stacks'' of candidates having astrophysical
characteristics in common.  A modest excess may be statistically
significant if it is not penalized for a large number of trials. Neutrons and photons arise from the same types of pion-producing interactions.  The photon path
length exceeds the path length for EeV neutron decay, so this study is
sensitive to sources in a larger volume than just the Galaxy.  

The absence of detectable point sources of EeV neutral particles does
not mean that the sources of EeV rays are extragalactic.  It
might be that EeV cosmic rays are produced by transient sources such
as gamma ray bursts or supernovae.  The Auger Observatory has been
collecting data only since 2004.  It is quite possible that it has not
been exposed to neutral particles emanating from any burst of cosmic-ray production.  Alternatively, it is conceivable that there are
continuous sources in the Galaxy which emit in jets and are relatively
few in number, and if so none of those jets are directed toward Earth.  The
protons would be almost isotropized by magnetic fields, but neutrons
and gamma rays would retain the jet directions and would not arrive here.
Another possibility is that the EeV protons originate in sources with
much lower optical depth for escaping than is typical of the
known TeV sources.  The production of neutrons and photons at the
source could be too meager to make a detectable flux at Earth.

The null results from this search for point sources of photons is
nevertheless interesting in light of the stringent upper limits on
cosmic-ray anisotropy at EeV energies \citep{Auger:2012an,Abreu:2012ybu, Abreu:2011ve}. EeV protons originating near the Galactic plane, whether
from transient sources or steady sources, are expected to cause an
anisotropy that exceeds the observational upper limit. Those
expectations are not free of assumptions about magnetic field
properties away from the Galactic disk, so the case against Galactic
EeV proton sources is by no means closed. However, evidence for such
sources remains absent, despite a sensitive search for any flux of
EeV photons.

\acknowledgments
\section*{Acknowledgements}
\begin{flushleft}
The successful installation, commissioning, and operation of the Pierre Auger Observatory would not have been possible without the strong commitment and effort from the technical and administrative staff in Malarg\"{u}e. 

We are very grateful to the following agencies and organizations for financial support: 
Comisi\'{o}n Nacional de Energ\'{\i}a At\'{o}mica, Fundaci\'{o}n Antorchas, Gobierno De La Provincia de Mendoza, Municipalidad de Malarg\"{u}e, NDM Holdings and Valle Las Le\~{n}as, in gratitude for their continuing cooperation over land access, Argentina; the Australian Research Council; Conselho Nacional de Desenvolvimento Cient\'{\i}fico e Tecnol\'{o}gico (CNPq), Financiadora de Estudos e Projetos (FINEP), Funda\c{c}\~{a}o de Amparo \`{a} Pesquisa do Estado de Rio de Janeiro (FAPERJ), S\~{a}o Paulo Research Foundation (FAPESP) Grants \# 2010/07359-6, \# 1999/05404-3, Minist\'{e}rio de Ci\^{e}ncia e Tecnologia (MCT), Brazil; MSMT-CR LG13007, 7AMB14AR005, CZ.1.05/2.1.00/03.0058 and the Czech Science Foundation grant 14-17501S, Czech Republic;  Centre de Calcul IN2P3/CNRS, Centre National de la Recherche Scientifique (CNRS), Conseil R\'{e}gional Ile-de-France, D\'{e}partement Physique Nucl\'{e}aire et Corpusculaire (PNC-IN2P3/CNRS), D\'{e}partement Sciences de l'Univers (SDU-INSU/CNRS), France; Bundesministerium f\"{u}r Bildung und Forschung (BMBF), Deutsche Forschungsgemeinschaft (DFG), Finanzministerium Baden-W\"{u}rttemberg, Helmholtz-Gemeinschaft Deutscher Forschungszentren (HGF), Ministerium f\"{u}r Wissenschaft und Forschung, Nordrhein Westfalen, Ministerium f\"{u}r Wissenschaft, Forschung und Kunst, Baden-W\"{u}rttemberg, Germany; Istituto Nazionale di Fisica Nucleare (INFN), Ministero dell'Istruzione, dell'Universit\`{a} e della Ricerca (MIUR), Gran Sasso Center for Astroparticle Physics (CFA), CETEMPS Center of Excellence, Italy; Consejo Nacional de Ciencia y Tecnolog\'{\i}a (CONACYT), Mexico; Ministerie van Onderwijs, Cultuur en Wetenschap, Nederlandse Organisatie voor Wetenschappelijk Onderzoek (NWO), Stichting voor Fundamenteel Onderzoek der Materie (FOM), Netherlands; National Centre for Research and Development, Grant Nos.ERA-NET-ASPERA/01/11 and ERA-NET-ASPERA/02/11, National Science Centre, Grant Nos. 2013/08/M/ST9/00322 and 2013/08/M/ST9/00728, Poland; Portuguese national funds and FEDER funds within COMPETE - Programa Operacional Factores de Competitividade through Funda\c{c}\~{a}o para a Ci\^{e}ncia e a Tecnologia, Portugal; Romanian Authority for Scientific Research ANCS, CNDI-UEFISCDI partnership projects nr.20/2012 and nr.194/2012, project nr.1/ASPERA2/2012 ERA-NET, PN-II-RU-PD-2011-3-0145-17, and PN-II-RU-PD-2011-3-0062, the Minister of National  Education, Programme for research - Space Technology and Advanced Research - STAR, project number 83/2013, Romania; Slovenian Research Agency, Slovenia; Comunidad de Madrid, FEDER funds, Ministerio de Educaci\'{o}n y Ciencia, Xunta de Galicia, Spain; The Leverhulme Foundation, Science and Technology Facilities Council, United Kingdom; Department of Energy, Contract No. DE-AC02-07CH11359, DE-FR02-04ER41300, and DE-FG02-99ER41107, National Science Foundation, Grant No. 0450696, The Grainger Foundation, USA; NAFOSTED, Vietnam; Marie Curie-IRSES/EPLANET, European Particle Physics Latin American Network, European Union 7th Framework Program, Grant No. PIRSES-2009-GA-246806; and UNESCO.
\end{flushleft}


\end{document}